# The Impact of Space Debris on Optical Astronomy

Anthony Mallama [1], Sergey Karpov [2] and Richard E. Cole [1]

[1] IAU Centre for the Protection of the Dark and Quiet Sky,
IAU-UAI Headquarters, 98-bis Blvd Arago, 75014,
Paris, France

[2] Institute of Physics, Czech Academy of Sciences,
CZ-182 21 Prague 8, Czech Republic

Correspondence: anthony.mallama@gmail.com

Photometric observations of space debris objects brighter than magnitudes 6.0, 7.0 and 8.0 are characterized. Those magnitudes pertain to different levels of interference with astronomy as defined by the International Astronomical Union. The densities of debris objects per square degree of sky and brighter than the magnitudes listed above are only 6, 13 and 24 parts per million, respectively. Furthermore, only about 1.4%, 2.5% and 3.8% of the 12,173 debris objects cataloged by NORAD exceed those magnitudes for even a small fraction of the time. So, debris trails have a minor impact on astronomical images and visual sightings are few. Characteristics of debris resulting from selected spacecraft collisions and rocket explosions are also quantified. This study is based on 13 million observations recorded by the MMT9 observatory.

## 1. Introduction

Bright spacecraft interfere with astronomical observations (Barentine et al. 2023) and spoil the aesthetic beauty of the night sky (Mallama and Young 2021). This problem has worsened with the launch of satellite megaconstellations.

Brightness statistics for all these spacecraft have been reported by Mallama and Cole (2025) and Mallama et al (2026) using magnitudes recorded by the Mini-MegaTORTORA (MMT9) robotic observatory (Karpov et al. 2016; Beskin et al. 2017) supplemented with visual observations. Those results agree closely with values for three models of Starlink satellites derived by Zhi et al (2024). Other recent studies including that by Longa-Peña et al (2026) report similar brightness characteristics.

This study expands our survey to include space debris. Some of these objects are remnants of satellite launches such as rocket upper stages, fairings and covers. Others are items resulting from intentional and accidental destruction of orbiting spacecraft as well as explosions of rocket upper stages.

Section 2 of this paper discusses three levels of brightness that interfere with astronomy. Section 3 describes the MMT9 system and its vast database of debris object observations. Section 4 examines the recorded magnitude distribution of these objects. Section 5 focuses on the year 2025 in order to assess the recent situation in this rapidly changing field. Section 6 describes debris from destructive events such as those mentioned above. Section 7 discusses the results reported here in the context of related studies. Section 8 summarizes our conclusions.

**2. Brightness limits for astronomy**

The International Astronomical Union (IAU, 2024) defined upper limits to brightness that minimize interference with observational research. For altitudes up to 550 km that limit is Johnson visual magnitude, $M_V$, 7.0. Equation 1 applies above that height,

$$M_V = 7.0 + 2.5 * \log10 ( \text{altitude} / 550 ) \quad \text{Equation 1}$$

where *altitude* refers to the satellite's height above sea level in km. The equation evaluates to $M_V$ = 8.0 for an altitude of 1,380 km which exceeds the height for most objects in low Earth orbit.

The IAU statement also recommends that spacecraft should not be visible to the unaided eye. Objects of visual magnitude 6.0 can be seen at locations where the sky is minimally affected by light pollution. The magnitude 6, 7 and 8 limits are used to assess the impact of debris objects on astronomy in this study.

**3. Observations**

Photometric measurements are recorded by the Mini-MegaTORTORA (MMT9) robotic observatory in Russia at 43.65N and 41.43E (Karpov et al. 2016; Beskin et al. 2017). The observatory has been operating since 2014 and its magnitude measurements are within 0.1 of the Johnson V-band as discussed by Mallama (2021).

MMT9 records satellite data in tracks across the sky at a frequency of 10 Hz. We extracted from an online database of MMT9 satellite observations (Karpov et al. 2016) the list of magnitudes for all objects in the NORAD catalog of orbiting bodies that contained 'DEB' in their names indicating that they are debris objects. This list contains 13.2 million magnitude measurements from 21,287 tracks corresponding to 1,098 objects.

The publicly available portion of the MMT9 database does not include objects associated with the launch of Russian satellites. So, those objects are not part of this study. However, we derived a correction factor based on the results of manually sampling every tenth debris object in the NORAD catalog and determining its country of origin. We found that 378 of the 1,217 sampled objects were associated with Russian launches, which indicates that 839 were from other countries. Therefore, a correction factor of 1.45 ( = 1 + 378 / 839) can be applied to compute total numbers of debris objects where needed.

**4. Observed brightness distribution**

The MMT9 data were sorted into magnitude bins and are plotted in Figure 1 to show the observed brightness distribution. The peak is at magnitude 8.5 and the decline at fainter magnitudes is a selection effect caused by instrument sensitivity. The decline at brighter magnitudes is real and is due to smaller numbers of large debris objects.

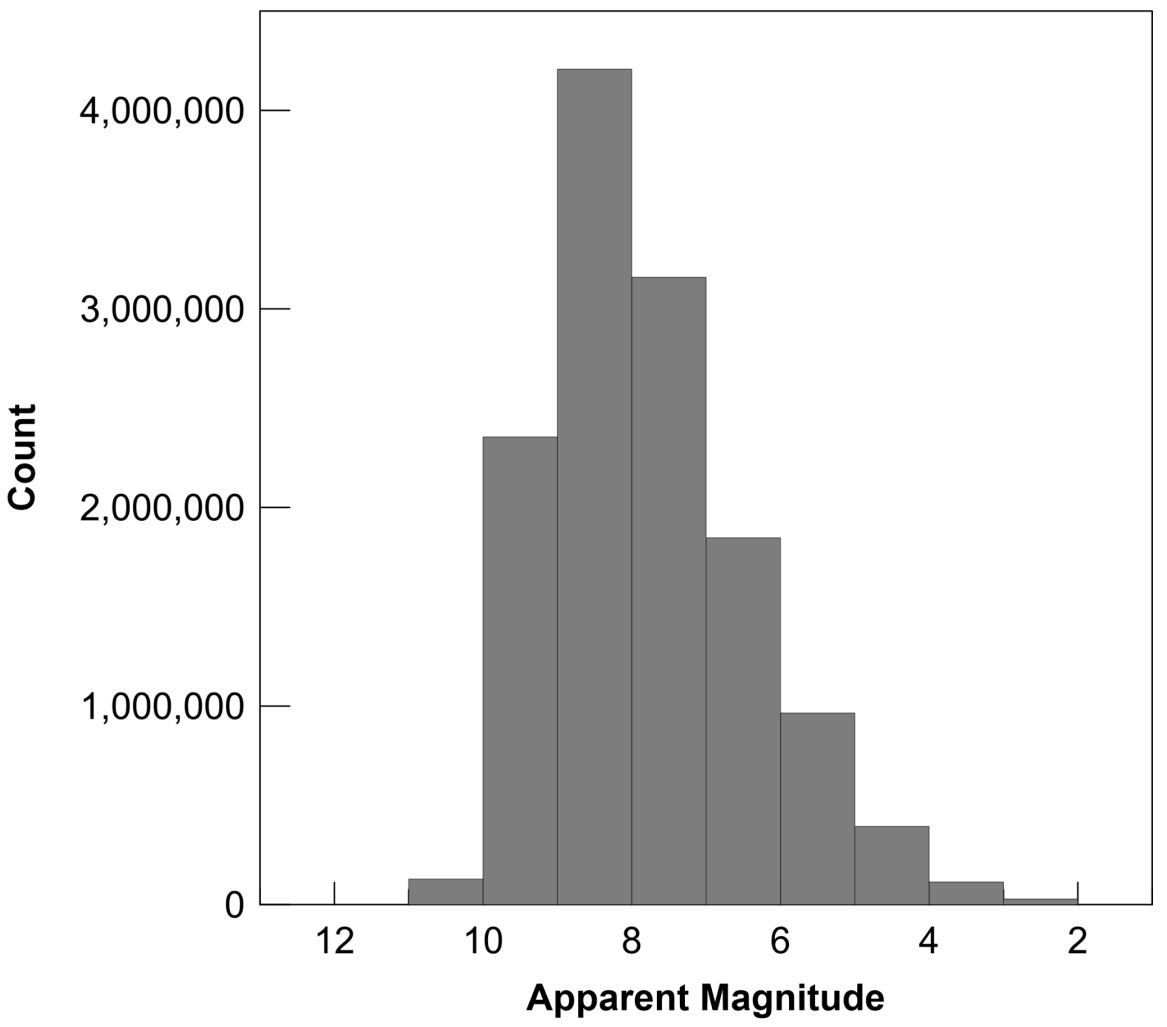

*Figure 1. The distribution of apparent magnitudes recorded for all debris objects and all years.*

Examination of light curves for many tracks of variable brightness objects indicates that MMT9 records them with very high reliability to magnitude 8, frequently to magnitude 9 and sometimes as faint as magnitudes 10 and 11. Since the trails of satellites fainter than magnitude 8 do not generally impact astronomical images, the data recorded by MMT9 is fully adequate for this research.

**5. The megaconstellation era**

The magnitude distribution described above applies to observations recorded by MMT9 since its inception in 2014. However, the population of objects in low Earth orbit has changed significantly during the era of megaconstellations. So, this section focuses on the year 2025 in order to address the recent impact of satellite debris on astronomical observing.

Counts of observations with magnitudes brighter than 8.0 are listed in Table 1. In order to quantify the impact of space debris, densities were computed for observations brighter than 6.0, 7.0 and 8.0 in units of objects per square degree of sky. The field of view for each MMT9 sensor is 99 square degrees, the total observing time for all 9 channels taken together during 2025 was 22.3 million seconds and observations were recorded at 10 Hz. Thus, the sky coverage for 2025 was 22.1 billion square-degrees. The number of debris observations brighter than magnitude 6.0, 7.0 and 8.0 are 96,968 195,241 and 369,896, respectively. When the correction for missing objects (described in Section 3) is applied, the numbers of observations become 140,604, 283,099 and 536,349. So, their densities are 6, 13 and 24 parts per million (ppm). These small numbers indicate that trails from debris objects are not seriously impacting optical astronomy, nor are they significantly interfering with aesthetic appreciation of the night sky.

The numbers of debris objects themselves are also worth noting. Those recorded one or more times brighter than magnitudes 6.0, 7.0 and 8.0 in 2025 were 119, 206 and 322, respectively. When the correction factor for the missing objects is applied, the numbers become 173, 299 and 467. In comparison to the 12,173 debris objects cataloged by NORAD the recorded objects represent just 1.4%, 2.5% and 3.8% of the total. This implies that few debris objects are bright enough to interfere with astronomy. Meanwhile, nearly all satellites in megaconstellations are brighter than magnitude 8.0 during the majority of their observed passes, and most are brighter than magnitudes 7.0 and 6.0 as well (see

Mallama and Cole 2025 and the figure shown here).

Table 1. Counts of observations of debris objects recorded in 2025 by magnitude.

| Bin center | Count | Cumulative |
|---|---|---|
| -1.5 | 22 | 22 |
| -0.5 | 55 | 77 |
| 0.5 | 62 | 139 |
| 1.5 | 205 | 344 |
| 2.5 | 1,372 | 1,716 |
| 3.5 | 12,226 | 13,942 |
| 4.5 | 24,549 | 38,491 |
| 5.5 | 58,477 | 96,968 |
| 6.5 | 98,273 | 195,241 |
| 7.5 | 174,655 | 369,896 |

MMT9 generally records multiple tracks for bright objects over the course of a year. For example, NORAD 51 (debris from Echo 1) has been tracked 182 times over the 12 years of MMT9 operation; an average of 15 times per year. So, the sampling of year 2025 data includes nearly all of the bright objects currently in orbit.

The brightest debris object at magnitude -2.30 is from NORAD 38345 (an H-IIA launch vehicle). The dimmest at 11.91 is from NORAD 21287 (a Delta rocket).

Another point that emphasizes the minor impact of debris objects is that MMT9 has only recorded 1,098 of them in 14 years of observing. That number increases to 1,592 when the correction factor for the missing objects is applied. The result is that ~87% have never been detected. So, they are practically always fainter than magnitude 9.0.

Finally, the debris objects launched before 2020.0 and after that date were investigated separately. We take 2020.0 to be the start of the full-scale megaconstellation era. Most satellites have been launched in large batches since that time. Figure 2 plots the total number (grey) of observations in magnitude bins along with those of pre-2020.0 objects (orange) and post 2020.0 objects (blue). Counts of the debris objects from earlier launches still exceed those for later ones in every magnitude bin even after six years of megaconstellation activity. This contrasts with the number of orbiting satellites launched after 2020.0 which already exceeds that for all earlier launches. Thus, constellation launches are generating fewer debris items per spacecraft than previous launches.

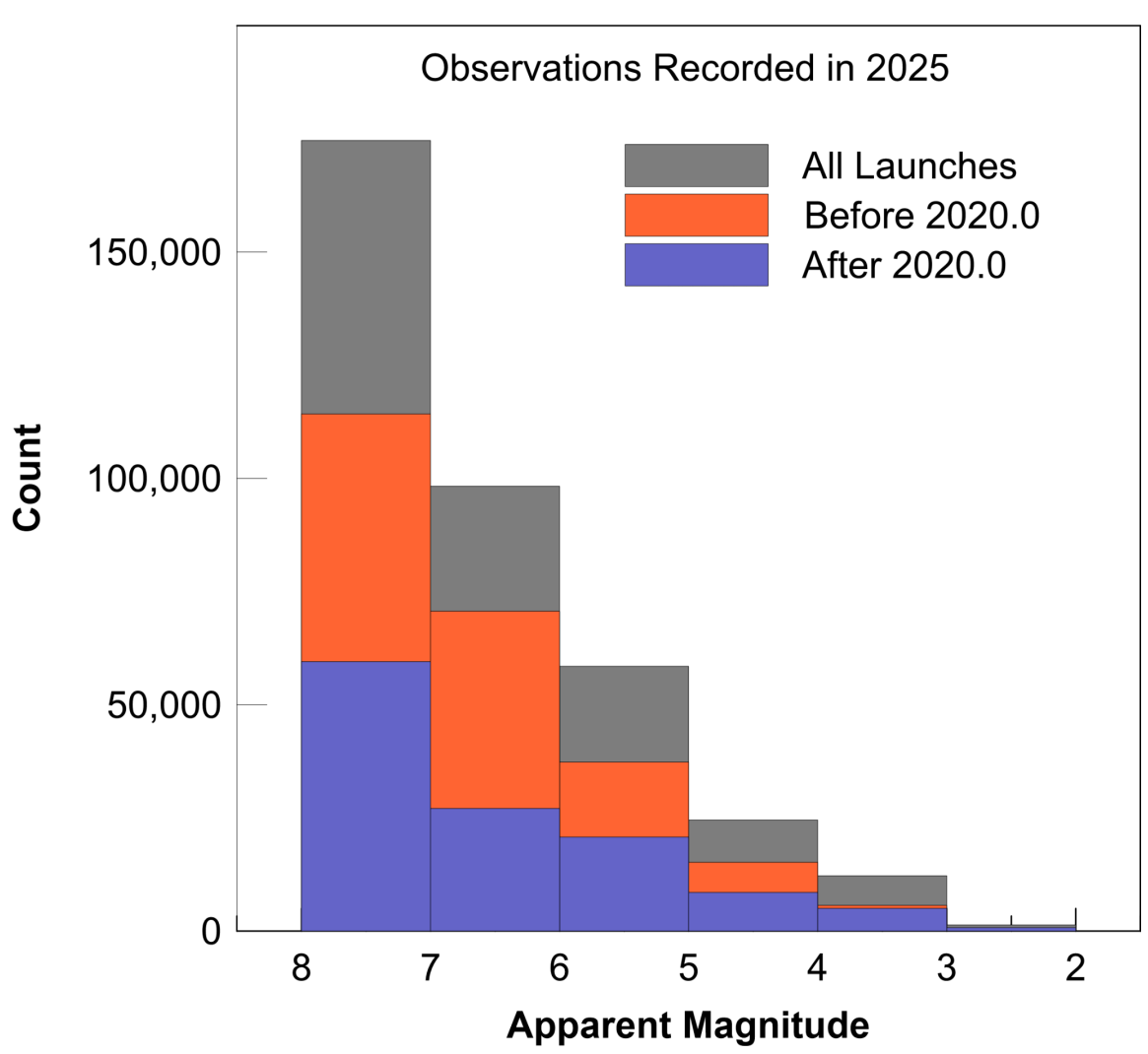

*Figure 2. Distribution of debris magnitudes brighter than 8.0 recorded during 2025. The bars all begin at zero; they are not stacked.*

**6. Debris from destructive events**

This section addresses several occurrences where the destruction of satellites and rockets in space produced many debris objects. A list of the top debris producing events, sorted by the number of resulting objects, summarizes the more extensive data compiled by J. McDowell on his Space Debris Clouds site.

First on the list is the anti-satellite test that destroyed the weather satellite Fengyun 1-C in 2007. An incoming missile, traveling at 8 km/s produced 3,549 pieces of debris. However, there are only 284 observations of two objects in the MMT9 data for 2025. So, the event is not seriously impacting astronomy at the present time.

The next two events on the list are an accidental collision of Kosmos 2551 with Iridium-33 in 2009 and an anti-satellite test which destroyed Kosmos 1408 in 2021. The two Kosmos satellites broke up into 1,716 and 1,562 pieces, respectively. Since the Kosmos spacecraft were Russian objects the MMT9 database does not contain publicly available observations.

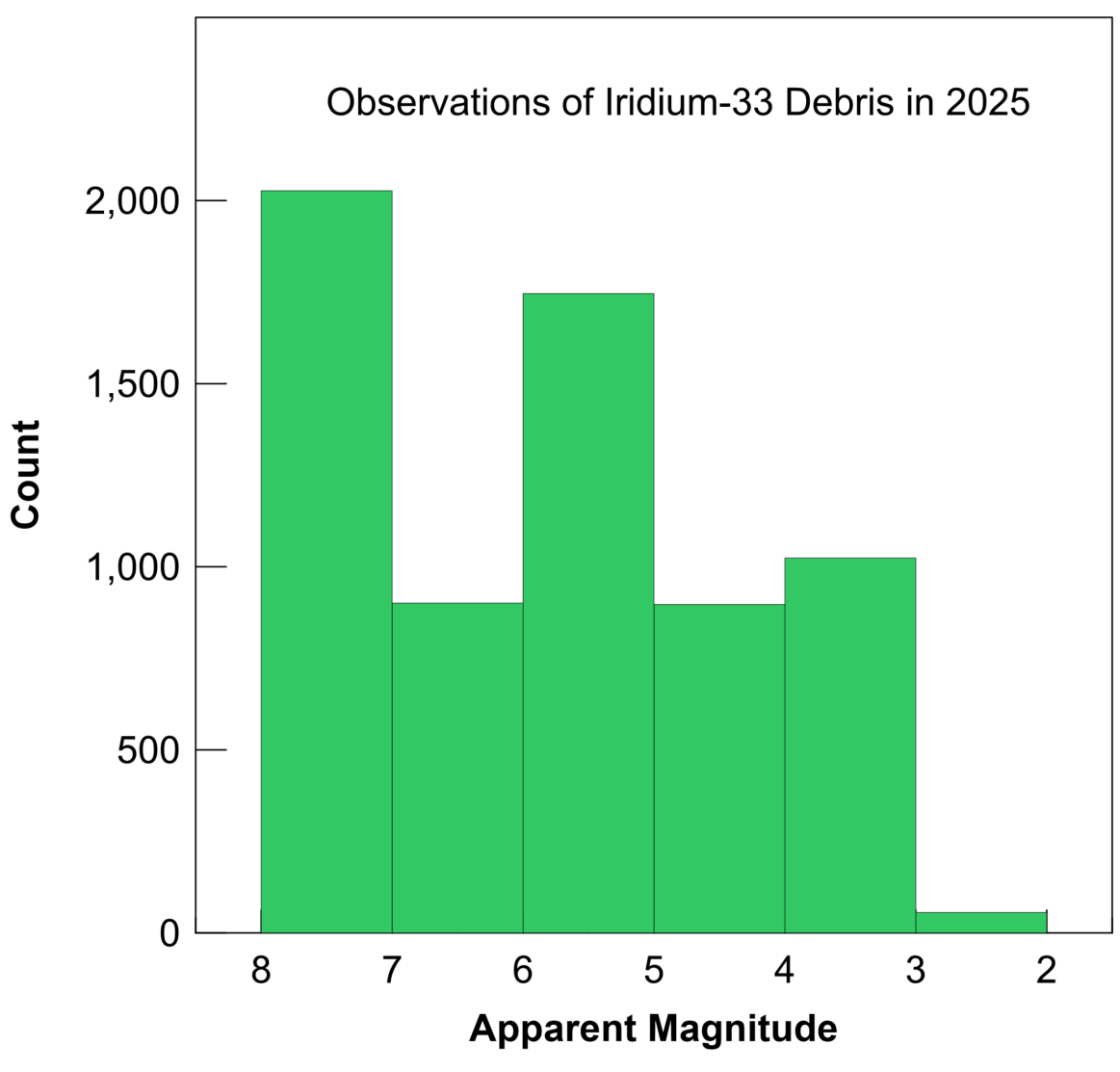

*Figure 3. Distribution of Iridium-33 debris magnitudes brighter than 8.0.*

The destruction of Iridium-33 itself generated 659 pieces of debris. The MMT9 data contains 8,955 observations of 10 objects recorded during 2025. The number of Iridium-33 observations is small compared to that for all debris objects and their brightness characteristics are unusual. Figure 3 shows that the distribution does not decrease rapidly for very bright magnitudes like the other groups of objects studied here. In fact, magnitudes brighter than 6.0 constitute about one-third of the total Iridium-33 observations.

The fourth and fifth events on the list both involved the break up of Long March 6A rocket upper stages. The debris objects are identified as CZ-6A DEB in the NORAD catalog and the two occurrences are distinguished by their years of launch which are 2022 and 2024. Each event produced about 800 pieces of debris.

The two CZ-6A events taken together account for about half of the bright observations of debris objects associated with all launches after 2020.0. The data recorded by MMT9 in 2025 is illustrated in Figure 4. Observations of debris from the 2024 launch are more numerous than those from 2022.

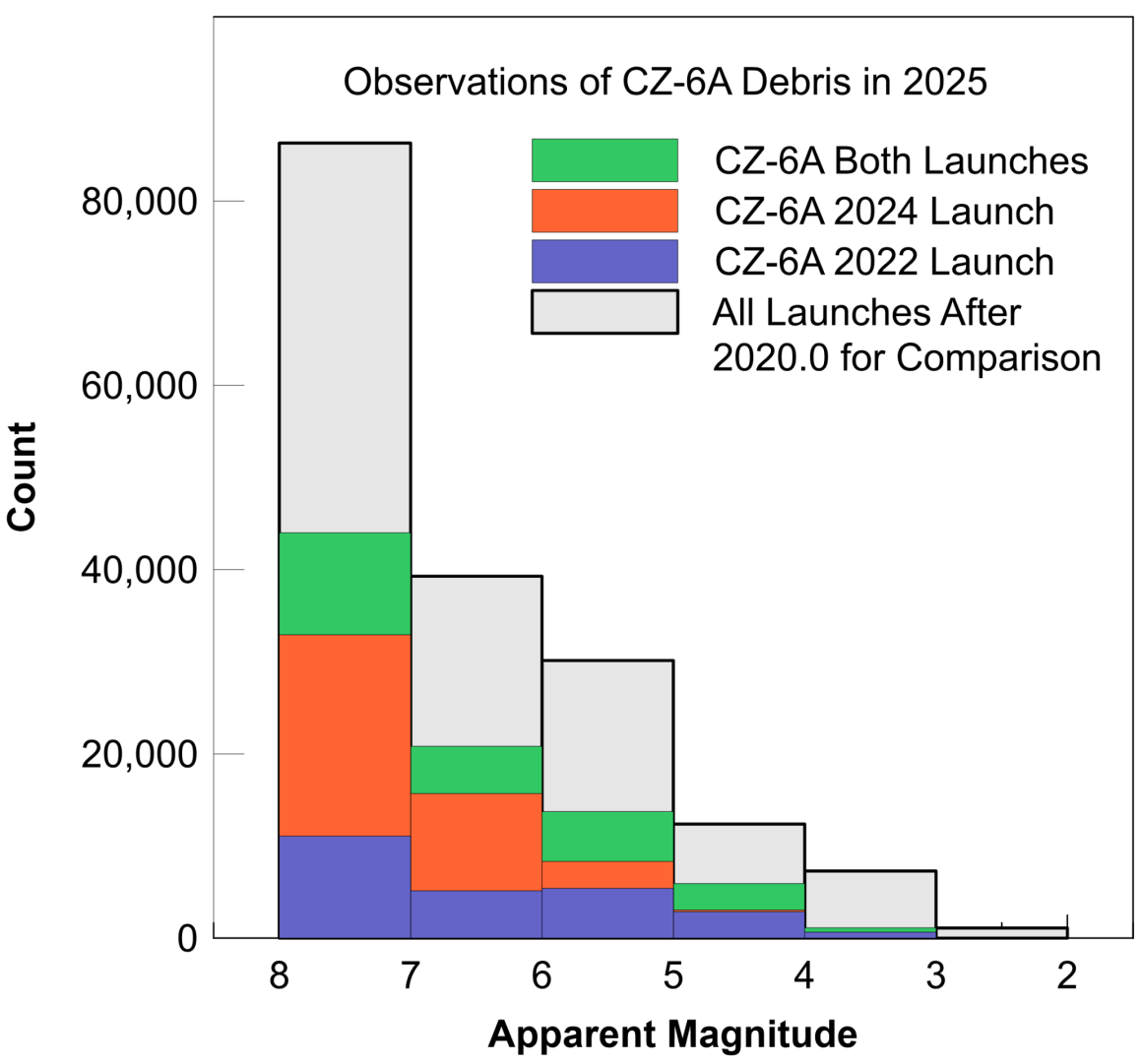

*Figure 4. Distribution of CZ-6A debris magnitudes for the launches of 2022 and 2024, and comparison with that for all launches after 2020.0. The correction factor for missing objects was applied to the data for 'All Launches'. The bars all begin at zero; they are not stacked.*

## 7. The context of this research

Rachith et al (2025) are performing research that is closely related to the work reported in this paper. Their observations from the VST/OmegaCAM archive are more sensitive than those studied here. The ongoing analysis aims to quantify the physical and rotational characteristics of debris.

Additionally, Hainaut (2026) performed a comprehensive study pertaining to the impact of large and bright satellite constellations due to satellite streaks, diffuse background, and scattered sky brightness. He noted that, "Bright satellites, such those from AST SpaceMobile, significantly impact saturating detectors even when their number is moderate." Meanwhile, the cumulative number of debris observations brighter than magnitude 5.0 listed in Table 1 is 38,491. After correcting by the missing objects factor, that number becomes 55,812. The corresponding density is only 3 ppm debris objects per square degree of sky.

Furthermore, removing debris trails from images requires additional analysis techniques and special processing.

Finally, S. Hellmich (private communication) pointed out that millions of uncatalogued debris particles are thought to exist. They can measurably increase sky brightness according to Wallner et al (2026) and Kocifaj et al (2025).

## 8. Conclusions

Trails from space debris have much less impact on astronomical images than do those from satellites. Likewise, debris is not significantly interfering with aesthetic appreciation of the night sky.

The densities of these objects brighter than magnitudes 6.0, 7.0 and 8.0 per square degree of sky are just 6, 13 and 24 ppm, respectively. Furthermore, only about 1.4%, 2.5% and 3.8% of the 12,173 debris objects cataloged by NORAD exceed the magnitudes listed above for even a small fraction of the time.

Observations brighter than magnitude 8.0 from debris objects associated with launches prior to 2020.0 (the beginning of the megaconstellation era) still exceed those for later launches even by 2025. This contrasts with the number of satellites that were launched after 2020.0 which already exceeds that for all earlier launches.

Launching batches of satellites together reduces the incidence of bright debris observations per satellite. Improved compliance with deorbit and passivation regulations and recommendations and the use of lower orbits leading to faster reentry are factors in this reduction.

Debris from two explosions of Long March 6A rockets accounts for about half of the observations that exceed magnitude 8.0 for launches after 2020.0. The accidental destruction of the Iridium-33 satellite produced debris objects with an excess of magnitudes brighter than 6.0.

**Data availability**

The observations studied in this article are available from the Mini-MegaTORTORA (MMT9) online database. Software used for the analysis is available from the corresponding author.

**Acknowledgements**
M. Dickinson of the IAU-CPS conducted a review of this study. Comments and suggestions from two anonymous reviewers helped to improve the paper.

The authors acknowledge the support of the International Astronomical Union (IAU) Centre for the Protection of the Dark and Quiet Sky (CPS, https://cps.iau.org). Any opinions, findings, and conclusions or recommendations expressed in this material are those of the author and do not necessarily reflect the views of the IAU, NSF NOIRLab, SKAO, ESO, or any host or member institution of the IAU CPS.